\documentclass[
  aps,
  prl,
  reprint,
  nofootinbib
]{revtex4-2}

\usepackage[T1]{fontenc}
\usepackage[utf8]{inputenc}
\usepackage{mathtools}
\usepackage{amssymb}
\usepackage{braket}
\usepackage{hyperref}
\usepackage{orcidlink}

\hypersetup{
  colorlinks=true,
  linkcolor=blue,
  citecolor=blue,
  urlcolor=blue,
  pdftitle={Answer Partitions and Oracle Access Determine Quantum Query Complexity},
  pdfauthor={Karl Svozil}
}

\begin{document}

\title{Answer Partitions and Oracle Access Determine Quantum Query Complexity}

\author{Karl Svozil \orcidlink{0000-0001-6554-2802}}
\email{karl.svozil@tuwien.ac.at}
\affiliation{Institute for Theoretical Physics, TU Wien,
Wiedner Hauptstra{\ss}e 8--10/136, 1040 Vienna, Austria}

\date{\today}

\begin{abstract}
An answer partition specifies which oracle instances share an output, while
query complexity also depends on how those instances are accessed. We formulate
exact one-query answer-partition resolution as block discrimination of a unitary
oracle family. Applied to Deutsch's problem with a controlled response unitary
\(V\), the criterion shows that one query suffices if and only if \(-1\) is an
eigenvalue of \(V\); cyclic addition in odd response dimension therefore raises
the exact quantum cost to two queries, matching the classical cost while
preserving the same computational-basis point values. Holding the standard
Boolean point oracle fixed, we further show that balanced answer partitions
with equal-sized cells can have different exact classical and quantum query
complexities. A complete exact classification is given for all 35 balanced
three-bit partitions, while the unresolved four-bit cases are reported
explicitly as numerical candidates. These results separate the combinatorial
specification of an answer from the oracle-dependent distinguishability that
determines query complexity.
\end{abstract}

\maketitle

An oracle problem pairs an answer partition with an access model for the hidden
instance. The partition groups instances into cells whose members share the
same required answer. The access model determines what information each query
makes available, classically or quantum mechanically, and how that information
can be processed. Quantum query speedup therefore belongs to the task--oracle
pair. We make this viewpoint precise and then test it on standard point
oracles, response-unitary variants, and small exhaustive families of balanced
tasks.

This viewpoint also sharpens the usual intuition about quantum parallelism. A
quantum query prepares one coherent state, and the algorithm succeeds when
interference makes the states associated with different answer cells
distinguishable. The useful resource is consequently the structure of the
oracle response---its phases, spectrum, accessible registers, and possible
environment---rather than simply the formal number of branches in a
superposition. Rewriting an oracle by fixed basis changes preserves this
discrimination, whereas changing its response representation, aggregating
information, or discarding a witness can alter it~\cite{WojcikChhajlany2006}.
The examples that follow develop these ideas through exact one-query
discrimination, parity, Bernstein--Vazirani, and a finite census of balanced
partitions.

Let \(\Omega\) be a finite instance set and let
\(g:\Omega\rightarrow J\) specify the required answer. The task induces the
\emph{answer partition} (also called here the kinematic partition)
\begin{equation}
 \Pi_g=\{C_j:j\in g(\Omega)\},\qquad C_j=g^{-1}(j).
 \label{eq:task-partition}
\end{equation}
This partition is indispensable bookkeeping, but it is not yet an access model
or a complexity statement.
In the query model the hidden instance labels a black-box transformation; the
algorithm is not handed a state \(\ket a\) on which the answer partition is a
pre-existing observable. The operational question is whether the specified
oracle and a fixed protocol can transfer the cell label to an accessible
output.

More explicitly, let \(\rho_a^{(T)}\) be the accessible state produced from
instance \(a\) after \(T\) queries. Exact resolution means that there is one
instance-independent measurement \(\{P_j\}\) satisfying
\begin{equation}
 P_jP_\ell=\delta_{j\ell}P_j,\,
 \sum_jP_j=I,\,
 \operatorname{Tr}(P_{g(a)}\rho_a^{(T)})=1.
 \label{eq:output-sectors}
\end{equation}
Consequently, supports belonging to different cells are orthogonal; no
orthogonality is required within a cell. The \(C_j\) themselves remain sets of
classical instances. They are not subspaces, and the answer sectors
\(P_j\mathcal H_{\rm out}\) arise only after an oracle and protocol have
encoded them. The partition therefore specifies a required zero pattern in
the output distinguishability matrix rather than a ready-made physical
observable.

We next characterize exact one-query resolution for a family of unitary
oracles \(\{U_a:a\in\Omega\}\). An arbitrary
one-query protocol may prepare a pure state \(\ket\Psi\) of the query system
\(Q\) and a reference \(R\), apply \(U_a\otimes I_R\), and then use an
instance-independent channel and measurement. Write
\begin{equation}
 \ket{\Psi_a}=(U_a\otimes I_R)\ket\Psi,
 \qquad
 \sigma=\operatorname{Tr}_R\ket\Psi\!\bra\Psi .
\end{equation}
The output Gram matrix before the final processing is
\begin{equation}
 G_{aa'}=\braket{\Psi_a|\Psi_{a'}}
 =\operatorname{Tr}\!\left(\sigma U_a^\dagger U_{a'}\right).
 \label{eq:gram}
\end{equation}

\paragraph{Proposition 1 (block discrimination).}
The partition \(\Pi_g\) is exactly resolvable with at most one use of
\(\{U_a\}\), allowing an arbitrary reference, if and only if there is a
density operator \(\sigma\) such that
\begin{equation}
 \operatorname{Tr}\!\left(\sigma U_a^\dagger U_{a'}\right)=0
 \quad\text{whenever}\quad g(a)\ne g(a').
 \label{eq:criterion}
\end{equation}
Indeed, exact discrimination requires the spans of the output states
associated with different cells to be orthogonal, which is equivalent to the
indicated pairwise inner products vanishing. Conversely, orthogonal cell
spans can be measured by their support projectors. Purification shows that
optimizing over \(\sigma\) already includes mixed probes and arbitrary
ancillary systems.

Equation~\eqref{eq:criterion}, with \(\sigma\succeq0\) and
\(\operatorname{Tr}\sigma=1\), is the direct block-discrimination form of
standard perfect unitary-discrimination conditions
\cite{Acin2001UnitaryDiscrimination,DuanFengYing2009}; it is not a new general
channel-discrimination theorem. Its role here is to connect an answer
partition to a specified oracle family. The Gram matrix need only be block
diagonal according to \(\Pi_g\); states within one cell may overlap or
coincide. Complete identification is the singleton-cell special case, where
\(G\) must be diagonal.

For bounded error, orthogonality is replaced by a common POVM with the required
success probability. Multiquery state conversion and function evaluation are
characterized by adversary methods~\cite{LeeEtAl2011StateConversion,
Reichardt2014SpanPrograms}.

Write \(Q_\epsilon(\Pi;\mathcal U)\) and
\(R_\epsilon(\Pi;\mathcal U_{\rm cl})\) for the optimal quantum and
classical worst-case numbers of calls needed with worst-case error at most
\(\epsilon\), under explicitly matched access conventions, and write
\(Q_E\equiv Q_0\) for exact quantum complexity.
Equation~\eqref{eq:criterion} characterizes exact resolvability with at most
one oracle use, namely \(Q_E(\Pi_g;\mathcal U)\le1\). For every nontrivial
partition, zero queries cannot resolve the task, so this is equivalent to
\(Q_E(\Pi_g;\mathcal U)=1\); the trivial one-cell partition instead has
\(Q_E(\Pi_g;\mathcal U)=0\). The criterion does not by itself establish an
advantage. Speedup additionally requires a gap between
\(Q_\epsilon\) and \(R_\epsilon\) for the same task and oracle interface.
Classically, a deterministic query protocol partitions \(\Omega\) by its
possible transcripts and succeeds exactly only when that transcript
partition refines \(\Pi_g\). A quantum protocol need not create such an
intermediate classical transcript; it changes the overlaps among the
hypothesis states and measures only after the required cells have separated.
The relevant resource is this rate of separation, not the number of formal
components in the probe superposition.

The criterion also distinguishes a change of representation from a change
of access. If
\begin{equation}
 V_a=A U_a B
 \label{eq:fixed-equivalence}
\end{equation}
for instance-independent unitaries \(A,B\), then
\(V_a^\dagger V_{a'}=B^\dagger U_a^\dagger U_{a'}B\); as the probe varies,
the attainable Gram matrices are unchanged. More generally, families that
exactly simulate one another with \(O(1)\) oracle calls have quantum query
complexities equal up to constant factors; an approximate simulation must
also charge its conversion error to \(\epsilon\). Preservation of a claimed
quantum--classical gap additionally requires corresponding constant-query
classical simulations. Exact one-query behavior can nevertheless change
when the response representation is not of the form
Eq.~\eqref{eq:fixed-equivalence}.

For example, Hadamard conjugation of the answer qubit turns
Eq.~\eqref{eq:xor-oracle} below into the full controlled-phase action
\(\ket{x,z}\mapsto(-1)^{z b_x}\ket{x,z}\), so the two forms are exactly
equivalent. The restricted sign oracle
\(P_b\ket x=(-1)^{b_x}\ket x\) is only a subroutine:
\(P_{b\oplus 1}=-P_b\), and the complementary truth tables differ by a
global phase. Supplying a coherent control for \(P_b\) restores the missing
relative phase, but coherent control is additional black-box access rather
than an automatic operation on an unknown unitary
\cite{AraujoFeixCostaBrukner2014}.

Deutsch's problem illustrates both the block criterion and this dependence.
Let \(b=(b_0,b_1)\) be the truth table of a Boolean function and use
\begin{equation}
 O_b\ket{x,y}=\ket{x,y\oplus b_x}.
 \label{eq:xor-oracle}
\end{equation}
With the answer qubit in \(\ket-\), phase kickback gives, up to the common
answer factor, the output states on the address register:
\begin{equation}
 \begin{aligned}
 \ket{\psi_{00}}&= \ket{+},&
 \ket{\psi_{11}}&=-\ket{+},\\
 \ket{\psi_{01}}&= \ket{-},&
 \ket{\psi_{10}}&=-\ket{-}.
 \end{aligned}
 \label{eq:deutsch-rays}
\end{equation}
Thus the constant and balanced cells occupy orthogonal rays, while the two
instances inside either cell are not distinguished. This is precisely a
block-diagonal, but not diagonal, Gram matrix and gives the familiar one
versus two exact-query separation~\cite{deutsch}.

The phenomenon is controlled by the response spectrum. Let \(V\) be any
unitary and consider the controlled-response family indexed by the same four
Boolean truth tables,
\begin{equation}
 U_b^{(V)}
 =\sum_{x=0}^1\ket x\!\bra x\otimes V^{b_x}.
 \label{eq:response-oracle}
\end{equation}
\paragraph{Theorem 1 (response spectrum).}
The Deutsch constant--balanced
partition is exactly one-query resolvable through
\(\{U_b^{(V)}\}\), with arbitrary probes and references, if and only if
\(-1\) is an eigenvalue of \(V\).

For necessity, write the probe as
\(\ket0\ket{\phi_0}+\ket1\ket{\phi_1}\), with \(\ket{\phi_x}\) on the
response and reference systems. Let
\(\rho_x=\operatorname{Tr}_R\ket{\phi_x}\!\bra{\phi_x}\succeq0\) be its
unnormalized response operator and put
\begin{equation}
 p_x=\operatorname{Tr}\rho_x,\qquad
 z_x=\operatorname{Tr}(\rho_x V),\qquad p_0+p_1=1.
 \label{eq:response-blocks}
\end{equation}
The four cross-cell orthogonality conditions include their complex
conjugates and reduce to
\begin{equation}
 z_0=z_0^*=-p_1,\qquad z_1=z_1^*=-p_0.
 \label{eq:response-constraints}
\end{equation}
Since \(|z_x|\le p_x\), these equations force
\(p_0=p_1=1/2\) and \(z_0=z_1=-1/2\). Equality in the unitary expectation
bound is possible only when the support of each nonzero \(\rho_x\) lies in
the \(-1\) eigenspace of \(V\). Conversely, a \(-1\) eigenvector provides
ordinary phase kickback and Eq.~\eqref{eq:deutsch-rays}.

Related qudit versions of Deutsch--Jozsa use higher-dimensional address and
response systems~\cite{Cereceda2004QuditDeutsch}. The theorem above instead
holds the four Deutsch instances fixed and varies the response unitary. It is
also distinct from prior-distribution ``useless-query'' bounds, which infer
quantum lower bounds from classical queries that reveal no information about a
property~\cite{MeyerPommersheim2011}.

\paragraph{Corollary 1 (cyclic response).}
For the elementary cyclic response \(X_d\ket y=\ket{y+1\bmod d}\), \(d\ge2\),
computational-basis queries return the same Boolean point value for every
\(d\) (by mapping \(y\mapsto y+b_x \pmod d\) with \(b_x\in\{0,1\}\)),
but \(-1\) occurs in the spectrum of \(X_d\) exactly when \(d\) is even.
Therefore
\begin{equation}
 Q_E^{(d)}=
 \begin{cases}1,&d\ \text{even},\\2,&d\ \text{odd},\end{cases}
 \qquad D^{(d)}=2.
 \label{eq:cyclic-response-cost}
\end{equation}
In particular, reversible qutrit addition removes Deutsch's exact quantum
advantage without changing its classical point values. This is a
constant-factor effect, not an asymptotic separation, but it demonstrates
that reversibility and blank-register behavior do not determine coherent
query power.

Parity provides a broader illustration of the same principle under standard
pointwise access. Let \(b\in\{0,1\}^N\) and
\(O_b\ket{i,y}=\ket{i,y\oplus b_i}\). Every pure probe, including a
reference system, can be written in the answer qubit's \(X\) basis as
\begin{equation}
 \ket\Psi=\sum_{i=1}^N\ket i
 \left(\ket+\ket{\alpha_i}+\ket-\ket{\beta_i}\right),
 \qquad p_i=\lVert\beta_i\rVert^2 .
 \label{eq:parity-probe}
\end{equation}
The overlap between the outputs for \(0^N\) and the odd string \(e_i\) is
\begin{equation}
 \braket{\Psi_{0^N}|\Psi_{e_i}}=1-2p_i.
 \label{eq:parity-overlap}
\end{equation}
Exact even--odd separation would require \(p_i=1/2\) for every \(i\), while
normalization gives \(\sum_i p_i\le1\). Hence one query is possible only
for \(N\le2\), with Deutsch as the limiting case. More generally, the
acceptance probability of a \(T\)-query algorithm is a polynomial of degree
at most \(2T\) in the input bits, whereas parity has exact degree \(N\).
Thus \(T\ge N/2\), and applying the Deutsch construction to disjoint pairs
gives the matching algorithm, so
\begin{equation}
 Q_E(\operatorname{Par}_N)=\lceil N/2\rceil,
 \qquad D(\operatorname{Par}_N)=N
 \label{eq:parity-complexity}
\end{equation}
for point-value access~\cite{bbcmw-01,Farhi-98}. A binary partition can
therefore remain query intensive. If instead one declares
\(\widetilde O_b\ket y=\ket{y\oplus\operatorname{Par}(b)}\) to be a single
oracle call, parity becomes one-query resolvable classically as well as
quantum mechanically. That does not defeat
Eq.~\eqref{eq:parity-complexity}; it replaces point access by an oracle that
already aggregates the desired answer.

The preceding examples point to a common picture: the answer partition specifies
the required outputs, whereas the query model determines the cost of separating
its cells. We test this picture exhaustively for small point-oracle problems. Let the hidden instance be
\(b\in\{0,1\}^N\), accessed through Eq.~\eqref{eq:xor-oracle} with the address
range enlarged to \(N\), and consider every binary answer map having
\(2^{N-1}\) instances in each cell. Exchanging the answer labels gives the same
unlabeled partition, so their number is
\begin{equation}
 M_N=\frac12\binom{2^N}{2^{N-1}}.
 \label{eq:balanced-count}
\end{equation}
Coordinate permutations, independent complements of coordinates, and answer
exchange preserve the point-oracle query cost. They reduce the \(M_3=35\)
partitions to six orbits and the \(M_4=6435\) partitions to 58. Orbit sizes are
restored when reporting the counts below.

For each partition we computed the optimal classic deterministic decision-tree depth
\(D\) and sought matching bounds on the exact quantum query complexity
\(Q_E\). One-query feasibility reduces, for the Boolean response, to rational
Gram constraints and was solved over the rationals. Multiquery feasibility was
formulated with the Gram-matrix semidefinite program of
Barnum, Saks, and Szegedy~\cite{BarnumSaksSzegedy2003}. Constructive upper
bounds were retained as adaptive trees whose nodes read either one bit or the
parity of two bits; the latter is one Deutsch query. Such trees are useful
certificates, but they are not a general characterization of exact quantum
algorithms~\cite{MontanaroJozsaMitchison2015}.

For \(N=3\) all lower and upper certificates meet exactly. Table~\ref{tab:n3}
shows every orbit, rather than only aggregate counts. Here
\(\mathrm{Sel}(c,u,v)=u\) for \(c=0\) and \(v\) for \(c=1\).

\begin{table}[t]
\caption{Balanced three-bit tasks up to coordinate permutations, input
complements, and answer exchange. The orbit sizes sum to 35.}
\label{tab:n3}
\begin{ruledtabular}
\begin{tabular}{lrrr}
Representative & Orbit & \(D\) & \(Q_E\)\\
\hline
\(x_1\) & 3 & 1 & 1\\
\(x_1\mathbin\oplus x_2\) & 3 & 2 & 1\\
\(\operatorname{Par}_3\) & 1 & 3 & 2\\
\(\operatorname{Maj}_3\) & 4 & 3 & 2\\
\(x_1\mathbin\oplus x_2x_3\) & 12 & 3 & 2\\
\(\mathrm{Sel}(x_1,x_2,x_3)\) & 12 & 2 & 2\\
\end{tabular}
\end{ruledtabular}
\end{table}

Thus 20 of the 35 balanced tasks admit an exact advantage: three by a factor
of two and seventeen by a factor of \(3/2\). The certificates also display the
conditional structure hidden by the bare partition. For example, three-bit
majority is decided by first obtaining \(b_1\oplus b_2\); if it is zero one
reads \(b_1\), and if it is one one reads \(b_3\). Likewise,
\(F=b_1\oplus b_2b_3\) is decided by first reading \(b_2\), followed by
\(b_1\) when \(b_2=0\) and by the Deutsch parity \(b_1\oplus b_3\) when
\(b_2=1\). Both real multilinear polynomials have degree three, so the
polynomial method excludes one query and proves these two-query constructions
optimal.

For \(N=4\), exact rational lower certificates and explicit trees settle 395
partitions: 4, 6, 48, 248, and 89 have \((D,Q_E)=(1,1),(2,1),(2,2),(3,2)\),
and \((4,2)\), respectively. Of these, 343 have an exact advantage. For the
remaining 43 orbit representatives, covering 6040 partitions, exact trees give
\(Q_E\le3\), while two SCS  (Splitting Conic Solver) runs at requested tolerances \(10^{-7}\) and
\(10^{-8}\) report the two-query SDP infeasible. The resulting assignments,
1536 at \((D,Q_E)=(3,3)\) and 4504 at
\((4,3)\), are therefore \emph{numerical candidates}, not exact lower bounds.
If accepted, they give candidate totals of 4847 speedup and 1588 no-speedup
partitions. Exact dual certificates are required before these totals can be
stated as theorems.

Numerical SDP studies of all Boolean functions through four bits were already
reported in Ref.~\cite{MontanaroJozsaMitchison2015}; subsequent structural work
on exact two-query algorithms also uses that four-bit numerical classification
\cite{YeYangHuang2024}. We cross-validated our orbit representatives against
the published NPN (Negation-Permutation-Negation)identifiers of Ref.~\cite{MontanaroJozsaMitchison2015}. After
removing dummy variables, its one- and two-query entries agree with all 15
representatives certified here. For the 43 remaining representatives, its
reported optimal two-query success probabilities lie between 0.900 and 0.986,
consistent with our infeasibility results but still numerical. This agreement
checks the enumeration and symmetry bookkeeping; it does not manufacture an
exact lower bound.

The narrower contribution of the census is therefore to reorganize balanced
functions as unlabeled answer partitions, quotient only by symmetries that
preserve point access, attach simple exact certificates where available, and
keep numerical status explicit. The observed proportions are finite-model
facts, not an asymptotic speedup criterion.

Bernstein--Vazirani provides a complementary access-model example. For
\(a,x\in\mathbb F_2^n\), let \(f_a(x)=a\cdot x \pmod2\). With ordinary phase
access, the corresponding character states are orthogonal, so the singleton
partition is resolved in one query~\cite{be-va}. We now add an inaccessible
address record while preserving every computational-basis value returned by
the oracle.

Choose \(S\subseteq\{1,\ldots,n\}\), \(|S|=k\), and let \(x_S\) denote
the corresponding substring. Consider the unitary dilation
\begin{equation}
 W_a^S\ket{x,y,e}
 =\ket{x,y\oplus a\cdot x,e\oplus x_S}.
 \label{eq:witness-unitary}
\end{equation}
Each call receives a fresh witness in \(\ket{0^k}\), which is discarded
afterward. The accessible oracle is consequently the channel
\begin{equation}
 \Phi_a^S(\rho)=\operatorname{Tr}_E\!\left[
 W_a^S(\rho\otimes\ket{0^k}\!\bra{0^k})(W_a^S)^\dagger
 \right].
 \label{eq:witness-channel}
\end{equation}
Basis queries still return \(a\cdot x\) exactly, while the witness records
selected address information and removes coherences between sectors with
different \(x_S\), as in standard which-way
interference~\cite{Englert1996}. Define the surviving answer partition
\begin{equation}
 \Pi_S=\{C_u:u\in\mathbb F_2^{n-k}\},\qquad
 C_u=\{a:a_{\bar S}=u\}.
 \label{eq:surviving-partition}
\end{equation}
The one-query statement below concerns this surviving partition \(\Pi_S\),
whose cells are indexed by \(a_{\bar S}\); it is therefore an
access-induced change of the task--oracle pair rather than a
same-partition comparison for the original singleton task. For the classical
comparison, let \(\mathcal O^{\mathrm{val}}=\{O_a^{\mathrm{val}}:a\in
\mathbb F_2^n\}\) denote the value-oracle family induced by
computational-basis preparation and readout of \(\Phi_a^S\): a query chooses
\(x\) and returns \(a\cdot x\). For \(k<n\) and worst-case error
\(\epsilon<1/2\), these explicitly specified quantum and classical interfaces
give
\begin{equation}
 Q_\epsilon^{\Phi}(\Pi_S)=1,\qquad
 R_\epsilon(\Pi_S;\mathcal O^{\mathrm{val}})=n-k.
 \label{eq:partition-speedup}
\end{equation}
For the quantum part, the usual Bernstein--Vazirani input and phase kickback,
followed by tracing out the witness, leave
\begin{equation}
 \rho_a^S=2^{-n}\!\sum_{x,x':\,x_S=x'_S}
 (-1)^{a\cdot(x\oplus x')}\ket x\!\bra{x'}.
 \label{eq:reduced-bv}
\end{equation}
Character orthogonality after the final Hadamards gives, up to a fixed
reordering of coordinates,
\begin{equation}
\begin{aligned}
 H^{\otimes n}\rho_a^S H^{\otimes n}
 &=\ket{a_{\bar S}}\!\bra{a_{\bar S}}\otimes\frac{I_S}{2^k},\\
 \Pr(z|a)&=2^{-k}\,\mathbf 1[z_{\bar S}=a_{\bar S}].
\end{aligned}
 \label{eq:leaky-distribution}
\end{equation}
Thus the one-query readout exactly resolves \(\Pi_S\). Zero queries cannot
resolve this nontrivial
partition \(\Pi_S\), because the output would be independent of its cell
label. The classical upper bound uses the basis queries outside \(S\). For the
lower bound, impose the favorable promise \(a_S=0\). Under the uniform
distribution on \(a_{\bar S}\), fewer than \(n-k\) returned linear equations
leave an affine solution space of positive dimension, so even a decoder given
the complete transcript succeeds with probability at most \(1/2\). The same
average bound holds after including the protocol's internal randomness, and
therefore some input has success at most \(1/2\). If the witness were
coherently available, the fixed coupling
\(\ket{x,e}\mapsto\ket{x,e\oplus x_S}\) could be undone without another
oracle call, restoring the ordinary Bernstein--Vazirani output. The
accessible oracle here is a channel rather than the globally unitary dilation,
highlighting the distinction between standard and garbage-bearing oracle
access in quantum query complexity~\cite{Aaronson2021OpenProblems}.

At \(k=0\), no address coordinate is recorded and
Eq.~\eqref{eq:leaky-distribution} is the ordinary Bernstein--Vazirani output
resolving the singleton partition, giving the \(n\)-versus-one
identification gap.
At \(k=n\), the stated interference readout is uniform and
\(\Pi_S\) has only one cell, although ordinary basis queries can still
recover \(a\). The model separates loss of the phase-interference route from
loss of the classical function values themselves.

We conclude by returning to the role of the answer partition. Equation
\eqref{eq:task-partition} specifies which hidden instances share an answer,
while the oracle and protocol provide the physical ingredients---the probe
state, oracle transformation, accessible output space, and measurement---that
resolve the corresponding answer classes. Together, these ingredients
determine how rapidly states from different cells separate. The
response-spectrum theorem and the contrast between point and aggregate parity
show how changing the access model can change this separation even when the
abstract task remains fixed. Conversely, the finite census holds the point
oracle fixed while varying the balanced target. Its exact three-bit
certificates show how quantum advantages can arise from oracle-accessible,
conditionally composed refinements, while the four-bit calculation clearly
separates exact certificates from numerical candidates.

Shor's and Grover's algorithms fit the same task--oracle perspective: Shor
exploits periodic structure in the oracle family accessible via the quantum Fourier transform,
while Grover exploits symmetry and coherent amplitude amplification via a matched phase oracle.
These examples reinforce that
quantum speedup depends on the structure of the oracle access, rather than on
the coarseness or cardinality of the answer partition alone.

Partitions are therefore indispensable as specifications but insufficient as
complexity criteria. Quantum advantage is not caused by partition coarseness
or by simultaneous access to imagined branch values. It occurs when
structure-matched interference separates the required answer cells in fewer
calls than any classical transcript under the same interface. Superlinear
exact separations for total Boolean functions are known
\cite{Ambainis2016SuperlinearExact}, but establishing any asymptotic speedup
requires a structured family of task--oracle pairs; percentages from a finite
census cannot substitute for that analysis.

\begin{acknowledgments}
The author thanks Noson S. Yanofsky for discussions that motivated this
work. The author used an OpenAI Codex language-model-based assistant for
structural editing, consistency checks, and implementation of the finite
partition census, and accepts responsibility for the final scientific
content. This research was funded in whole or in part by the
Austrian Science Fund (FWF), grant
\href{https://doi.org/10.55776/PIN5424624}{PIN5424624}. The author
acknowledges TU Wien Bibliothek for support through its Open Access Funding
Programme.
\end{acknowledgments}

\noindent\textit{Data availability.---}The source code, exact certificates,
numerical witnesses, generated tables, and the NPN cross-validation underlying
the finite census are available at
\href{https://svozil.github.io/publications/2026-qc-partition-census-supplement.zip}{https://svozil.github.io/publications/2026-qc-partition-census-supplement.zip}.

\bibliography{svozil}

\end{document}